\documentclass[11pt]{llncs}

\begin{document}
\title{Colliding Message Pairs for 23 and 24-step SHA-512}
\author{Somitra Kumar Sanadhya\thanks{This author is supported by the
Ministry of Information Technology, Govt. of India.} and Palash Sarkar}
\institute{Applied Statistics Unit,\\ Indian Statistical Institute,\\
203, B.T. Road, Kolkata,\\India 700108.\\ somitra\_r@isical.ac.in,
palash@isical.ac.in}
\maketitle
\begin{center}
1$^{\mbox{st}}$ September, 2008
\end{center}

\begin{abstract}
Recently, Indesteege et al.~\cite{eprint/report131} had described attacks against 23 and
24-step SHA-512 at SAC '08. Their attacks are based on the differential
path by Nikoli\'{c} and Biryukov~\cite{fse/NikolicB08}. The reported
complexities are $2^{44.9}$ and $2^{53}$ calls to the respective step
reduced SHA-512
hash function. They provided colliding message pairs for 23-step SHA-512 but
did not provide a colliding message pair for 24-step SHA-512. In this note
we provide a colliding message pair for 23-step SHA-512 and the first
colliding message pair for 24-step SHA-512. Our attacks use the 
differential path first described by Sanadhya and Sarkar at ACISP
'08~\cite{acisp/SanadhyaS08}. The complexities of our attacks are
$2^{16.5}$ and $2^{34.5}$
calls to the respective step reduced SHA-512 hash function.
Complete details of the attacks will be provided in 
an extended version of this note.
\end{abstract}

\section{Colliding Message Pairs}

In~\cite{eprint/generalDP}, 23 and 24-step SHA-256 attacks are described. Similar attacks
will also work for 23 and 24-step SHA-512. Complete details of these
attacks will be provided later.  For notation see~\cite{eprint/generalDP}.

A set of suitable values of 
$\delta_2$, $\alpha$, $\lambda$, $\mu$ and $\gamma$ for the 23-step SHA-512 collision is the following.
$\delta_2 = \texttt{0x600000000237}$, $\alpha
= \texttt{0x7201b90f9f8df85e}$,  $\lambda = \texttt{0x3e000007ffdc9}$, $\mu
= \texttt{0x43fffff800001}$ and $\gamma = \texttt{0x1}$. 

Values of the constants for 24-step SHA-512 collision is the following.
$\delta_1 = \texttt{0x200000000008}$, $\delta_2 = \texttt{0x600000000237}$, 
$\alpha =$ $\texttt{0x7201b90f9f8df85e}$, $\lambda = \texttt{0x3e000007ffdc9}$, 
$\mu = \texttt{0x45fffff800009}$, $\gamma = \texttt{0x1}$.
The colliding message pairs are provided in Table~\ref{tbl:23step_512}
and Table~\ref{tbl:24step_512} next.

\begin{table}
\caption{Colliding message pair for 23-step SHA-512 with standard IV. 
\label{tbl:23step_512}}
\begin{center}
\begin{tabular}{|c|c|r|r|r|r|} \hline 
$W_1$ & 0-3 & \texttt{b9fa6fc4729ca55c} & \texttt{8718310e1b3590e1} &
\texttt{1d3d530cb075b721} & \texttt{99166b30ecbdd705} \\ \cline{3-6} 
 & 4-7 & \texttt{27ed55b66c090b62} & \texttt{754b2163ff6feec5} &
\texttt{6685f40fd8ab08f8} & \texttt{590c1c0522f6fdfd} \\ \cline{3-6} 
 & 8-11 & \texttt{b947bb4013b688c1} & \texttt{d9d72ca8ab1cac04} &
\texttt{69d0e120220d4edc} & \texttt{30a2e93aeef24e3f} \\ \cline{3-6} 
 & 12-15 & \texttt{84e76299718478b9} & \texttt{f11ae711647763e5} &
\texttt{d621d2687946e862} & \texttt{0ee57069123ecc8b} \\ \hline 
$W_2$ & 0-3 & \texttt{b9fa6fc4729ca55c} & \texttt{8718310e1b3590e1} &
\texttt{1d3d530cb075b721} & \texttt{99166b30ecbdd705} \\ \cline{3-6} 
 & 4-7 & \texttt{27ed55b66c090b62} & \texttt{754b2163ff6feec5} &
\texttt{6685f40fd8ab08f8} & \texttt{590c1c0522f6fdfd} \\ \cline{3-6} 
 & 8-11 & \texttt{b947bb4013b688c2} & \texttt{d9d72ca8ab1cac03} &
\texttt{69d0e120220d4edc} & \texttt{30a3493aeef25076} \\ \cline{3-6} 
 & 12-15 & \texttt{84e76299718478b9} & \texttt{f11ae711647763e5} &
\texttt{d621d2687946e862} & \texttt{0ee57069123ecc8b} \\ \hline 
\end{tabular}
\end{center}
\end{table}

\begin{table}
\caption{Colliding message pair for 24-step SHA-512 with standard IV. 
\label{tbl:24step_512}}
\begin{center}
\begin{tabular}{|c|c|r|r|r|r|} \hline 
$W_1$ & 0-3 & \texttt{dedb689cfc766965} & \texttt{c7b8e064ff720f7c} &
\texttt{c136883560348c9c} & \texttt{3747df7d0cf47678} \\ \cline{3-6} 
 & 4-7 & \texttt{855e17555cfedc5f} & \texttt{88566babccaa63e9} &
\texttt{5dda9777938b73cd} & \texttt{b17b00574a4e4216} \\ \cline{3-6} 
 & 8-11 & \texttt{86f3ff48fd12ea19} & \texttt{cd15c6f8d6da38ce} &
\texttt{5e2c6b7b0411e70b} & \texttt{36ed67e93a794e66} \\ \cline{3-6} 
 & 12-15 & \texttt{1b65e96b02767821} & \texttt{04d0950089db6c68} &
\texttt{5bc9b9673e38eff3} & \texttt{b05d879ad024d3fa} \\ \hline 
$W_2$ & 0-3 & \texttt{dedb689cfc766965} & \texttt{c7b8e064ff720f7c} &
\texttt{c136883560348c9c} & \texttt{3747df7d0cf47678} \\ \cline{3-6} 
 & 4-7 & \texttt{855e17555cfedc5f} & \texttt{88566babccaa63e9} &
\texttt{5dda9777938b73cd} & \texttt{b17b00574a4e4216} \\ \cline{3-6} 
 & 8-11 & \texttt{86f3ff48fd12ea19} & \texttt{cd15c6f8d6da38ce} &
\texttt{5e2c6b7b0411e70c} & \texttt{36ed67e93a794e65} \\ \cline{3-6} 
 & 12-15 & \texttt{1b66096b02767829} & \texttt{04d0f50089db6e9f} &
\texttt{5bc9b9673e38eff3} & \texttt{b05d879ad024d3fa} \\ \hline 
\end{tabular}
\end{center}
\end{table}

\end{document}